\documentclass[12pt]{article}  
\usepackage{amsmath, amssymb}  
\usepackage{graphicx}  
\usepackage{xcolor}  

\title{Investigation of $\beta$-decay properties of neutron-rich Cerium isotopes}  

\author{  
	Jameel-Un Nabi$^{1,2}$,  
	Asim Ullah$^{2}$\thanks{Corresponding author: asimullah844@gmail.com},  
	Zeeshan Khan$^{1}$  
}  

\date{}  % Remove if you want the default date  

\begin{document}  
	\maketitle  
	
	\begin{center}  
		\textit{$^{1}$University of Wah, Quaid Avenue, Wah Cantt 47040, Punjab, Pakistan} \\  
		\textit{$^{2}$Faculty of Engineering Sciences, GIK Institute of Engineering Sciences and Technology, Topi 23640, Khyber Pakhtunkhwa, Pakistan}  
	\end{center}  
	
	\begin{abstract}  
	{Reliable and precise knowledge of the $\beta$-decay properties of neutron-rich nuclei is important for a better understanding of the $r$-process. We report the computation of $\beta$-decay properties of neutron-rich Cerium isotopes calculated within the proton-neutron quasiparticle random phase approximation (pn-QRPA)  approach. A total of 34 isotopes of Ce in the mass range 120 $\le$ A $\le$ 157 were considered in our calculation. Pairing gaps are recognized amongst the key parameters in the pn-QRPA model to compute Gamow-Teller (GT) transitions. We employed two different values of the pairing gaps obtained from two different empirical formulae in our computation. The GT strength distributions changed considerably with change in the pairing gap values. This in turn resulted in contrasting centroid and total strength values of the GT  distributions and led to differences in calculated half-lives using the two schemes. The traditional pairing gaps resulted in siginificant fragmentation of GT strength. However, the pairing gaps, calculated employing the formula based on separation energies of neutron and proton, led to computed half-lives in better agreement with the measured data.} 
%%%%%%%%%%%%%%%%%%%%%%%%%%%%%%%%%%%%%%%%

\end{abstract}
%
% Uncomment for keywords
\vspace{10pt}  
\noindent \textbf{Keywords:} $\beta$-decay half-lives, Gamow-Teller transitions, Branching ratios, Pairing gaps, pn-QRPA theory, Centroid, Total GT strength  
\section{Introduction}
The structure of a nucleus within the core of a massive star plays a crucial role in the process of a supernova explosion and determines the roadmap for evolution of stars \cite{Bet79}. According to numerical simulations, the stellar evolution is primarily determined by the temporal variation of the lepton fraction (Y$_e$), which is governed by the weak interaction (especially electron capture and $\beta$-decay). To explore the mechanism of supernova (both Type-Ia and Type-II) explosions, the magnitude of the electron capture rate is the most critical factor \cite{Ful80,Ful82a,Lan03}. In a Type-II supernovae, once the mass of the Fe core exceeds the Chandrasekhar mass limit, the degenerate electron gas pressure can no longer withstand gravitational force, and the core begins to collapse. Electron capture (\textit{ec}), on the one hand, lowers the lepton fraction, which in turn reduces the electron degenerate pressure. On the other hand, the \textit{ec} (and also the $\beta$-decay) results in the production of  (anti)neutrinos, which channel the energy away and may accelerate the collapse. A Type-Ia supernova is considered to be the result of a thermonuclear explosion on an accreting white dwarf, and its collapse is thought to be the result of the general relativistic effect. However, the \textit{ec} process is believed to be responsible for the abundance of certain iron isotopes in Type-Ia supernovae \cite{Lan03}.\\ 
Since these weak interaction processes are crucial for the evolution of supernovae and massive stars, the study of these processes has been the focus
of nuclear astrophysics. The calculations of  weak interaction rates under stellar conditions rely heavily on reliable computation of the ground and excited states Gamow-Teller response \cite{Bet79}. The Gamow-Teller (GT) excitation represents a very basic spin-isospin ($\sigma$$\tau$) nuclear response \cite{Ost92}. Owing to low core-densities ($\sim$ 10$^{10}$ g/cm$^3$) and temperatures (0.3 - 0.8 MeV) at start of core-collapse, the nuclear Q-value and chemical potential of electrons have comparable magnitudes. In such a scenario, the \textit{ec} rates are highly dependent on the comprehensive details of the distribution of GT strength. The centroid and total GT strength values control the \textit{ec} rates when the chemical potential surpasses the Q-value at relatively high core-densities. That is why, a thorough understanding of the GT distributions is required for calculating stellar rates and $\beta$-decay half-lives accurately.\\ Various charge-exchange (CE) reactions, such as (\textit{p}, \textit{n}) \cite{And91}, (\textit{n}, \textit{p}) \cite{El-94}, ($^3He$, \textit{t}) \cite{Fuj99} and (\textit{t}, $^3He$) \cite{Col06}, may be employed to extract the GT transitions experimentally. Modeling and simulation of core-collapse supernovae require the GT strength profiles of thousands of unstable nuclides. A microscopic nuclear model is needed to determine GT strength distributions (ground plus excited states) of hundreds of nuclei and related $\beta$-decay half-lives in excellent accordance with experimental data. \\  
To explore the $\beta$-decay properties and get a better knowledge of the stellar dynamics, numerous attempts have been made previously. {The calculations of Takahashi et al. \cite{Tak78} based on gross theory, QRPA based computations (e.g., \cite{Hir93,Nab99,Esc10,Wan16}) and shell model calculations (e.g., \cite{Mar99}), are noticeable mentions.} The gross theory computations failed to give structural information of each nucleus. Instead, they estimate the $\beta$-decay properties using a statistical approach. Among the microscopic techniques, the proton-neutron quasiparticle random phase approximation (pn-QRPA) and the shell model (SM) are extensively utilised for calculating half-lives of the $\beta$-decay and stellar weak rates. The SM, on the other hand, incorporates the contribution of excited state GT strength distributions at high temperatures using the Brink-Axel hypothesis \cite{Bri58}. The exited state GT transitions can be computed using the pn-QRPA model without using the Brink's hypothesis.\\
Since reliable estimates of $\beta$-decay half-lives of neutron-rich nuclei are necessary to better understand the mechanism of supernova explosion and the associated nucleosynthesis processes, particularly the $r$-process, this work focuses on the computation of $\beta$-decay half-lives and branching ratios of the GT strength distributions of the neutron-rich Cerium (Ce) isotopes via the pn-QRPA approach. {The chosen pn-QRPA model  deals with a simple pairing plus quadrupole nuclear Hamiltonian. It is to be noted that pn-QRPA model with realistic nucleon-nucleon interactions have been used in the past with great success. In the QRPA calculation~\cite{Mut89}, the authors used the G-matrix of the Paris potential. The QRPA, with realistic NN potential is used by the  T\"{u}bingen, Bratislava and Jyv\"{a}skyl\"{a} groups (e.g., \cite{Rod06}) and more recently by Jokiniemi and collaborators~\cite{Lot21}. The later two references used CD-Bonn, Nijmegen and Argonne parametrizations. The current pairing plus quadrupole Hamiltonian incorporates both particle-hole and particle-particle GT forces of separable form. This assumption of separable forces transforms the RPA matrix equation to an algebraic equation of fourth order. This, in turn, makes it possible to perform calculations for any arbitrary heavy nucleus due to availability of a large model space with up to seven major shells in the current model. The wavefunctions of the eigenstates are calculated accordingly. This feature of the current model has the added advantage of saving computation time in comparison with the full diagonalization of the non-Hermitian matrix of a very large dimension. Yet the compromised Hamiltonian gives a very decent comparison of $\beta$-decay half-lives with the measured data \cite{Hir93,Sta90}.  It is argued that the rate calculations do not depend significantly on the choice of the realistic NN interaction~\cite{Rod03}. A realistic NN potential is, nonetheless, likely to bring a marked improvement over a pairing plus quadrupole potential and may be taken as a future assignment.} The solution of the nuclear Hamiltonian complicates once transition is made from a schematic separable potential to a realistic NN potential. The potential would no more be separable and the advantage of solving an algebraic equation would be lost. We refer to \cite{Mut89} for solution of the pn-QRPA matrix equation using a realistic NN potential. Full diagonalization of a non-Hermitian matrix is a formidable task. The calculations are likely to be performed in a truncated model space. For further insight on using the pn-QRPA formalism for calculation of nuclear matrix elements using a realsitic NN potential we refer to \cite{Hir94}.  \\
{The lanthanide elements (Z = 57-71) are important in astrophysics in relation to nucleosynthesis and star formation considerations. Among all the lanthanides, Cerium is the most abundant in Ap Stars \cite{Cow76}. Because of its high abundance, there is a need to compute weak-interaction mediated rates of Cerium isotopes. From an astrophysical viewpoint, nuclei with A$\sim$100 have substantial importance. The r-process time scale and abundance peaks of isotopes at mass numbers A$\sim$80, 130 and 195 are determined by $\beta$-decay half-lives \cite{Cow91,Arn07,Mum14,Lan03,Kap11,Mum16,Mol03}. Consequently, half-lives of $\beta$ transitions of exotic neutron-rich nuclei are amongst the crucial nuclear inputs to study the r-process. }
For this project, a total of 34 isotopes of Ce in the mass range 120 $\le$ A $\le$ 157 were considered. The pairing gap between the nucleons is one of the important model parameters employed in the pn-QRPA calculation. We further investigate how the calculated $\beta$-decay properties vary by changing the pairing gap values in this work. \\
The following is the outline of the paper. The formalism employed in our calculation is briefly described in Section 2. The third section discusses our findings. The last section includes a summary and conclusion.    
%%%%%%%%%%%%%%%%%%%%%%%%%%%%%%%
\section{Theoretical Formalism}
The pn-QRPA approach has been used to compute the $\beta$-decay half-lives and branching ratios of the GT strength distributions of the Ce isotopes. Below we present a short necessary formalism of the model:\\
The Hamiltonian of the pn-QRPA model is given as:
\begin{equation} \label{H}
	H^{QRPA} = H^{sp} + V^{pair} + V^{pp}_{GT} + V^{ph}_{GT},
\end{equation}
where $H^{sp}$ stands for the single-particle Hamiltonian, $V_{GT}^{ph}$
and $V_{GT}^{pp}$ represent the particle-hole (\textit{ph}) and particle-particle (\textit{pp}) GT
forces, respectively. The last term $V^{pair}$ denotes the pairing
force which was computed under the assumption of the BCS approximation. The single-particle energies and wavefunction were computed using the Nilsson model \cite{Nil55}, which included nuclear deformation. $\hbar\omega=41A^{1/3}$ was used to calculate the oscillator constant for nucleons. The Nilson-potential
parameter were adopted from Ref. \cite{ragnarson1984}. $Q$-values were adopted from the recent compilation of Ref. \cite{Aud21}. The values of deformation parameter ($\beta_2$) were taken from Ref.~\cite{Mol16}. \\
The spherical nucleon basis ($c^{\dagger}_{jm}$, $c_{jm}$ with \textit{j} denotes the total angular momentum with \textit{m} as \textit{z}-component) was changed to the deformed basis ($d^{\dagger}_{m\alpha}$, $d_{m\alpha}$) employing the following transformation equation
\begin{equation}\label{df}
	d^{\dagger}_{m\alpha}=\Sigma_{j}D^{m\alpha}_{j}c^{\dagger}_{jm},
\end{equation}
where $c^{\dagger}$ ($d^{\dagger}$) is the particle creation operators in the spherical basis (deformed basis). The Nilsson Hamiltonian was diagonalized to get the matrices ($D^{m\alpha}_{j}$). We do separate BCS calculations for the neutron and proton systems. A pairing force of constant strength G ($G_n$ and $G_p$ for neutrons and protons, respectively), is given by
\begin{eqnarray}\label{pr}
	V_{pair}=-G\sum_{jmj^{'}m^{'}}(-1)^{l+j-m}c^{\dagger}_{jm}c^{\dagger}_{j-m}\\ \nonumber
	~~~~~~~~~~~~~~~~~(-1)^{j^{'}+l^{'}-m^{'}} c_{j^{'}-m^{'}}c_{j^{'}m^{'}},
\end{eqnarray}
with sum was limited to $m$. We introduced a quasiparticle basis $(a^{\dagger}_{m\alpha}, a_{m\alpha})$ from the Bogoliubov transformation
\begin{equation}\label{qbas}
	a^{\dagger}_{m\alpha}=u_{m\alpha}d^{\dagger}_{m\alpha}-v_{m\alpha}d_{\bar{m}\alpha}
\end{equation}
\begin{equation}
	a^{\dagger}_{\bar{m}\alpha}=u_{m\alpha}d^{\dagger}_{\bar{m}\alpha}+v_{m\alpha}d_{m\alpha},
\end{equation}
where $\bar{m}$ is the time-reversed states of $m$ and $a$ ($a^{\dagger}$) stands for the quasiparticle (q.p.) annihilation (creation) operator, which appears in the RPA equation later. The occupation amplitudes ($v_{m\alpha}$ and $u_{m\alpha}$) are calculated within BCS approximation (subject to $v^{2}_{m\alpha}$ + $u^{2}_{m\alpha}$ = 1).\\
The GT transitions in pn-QRPA approach are expressed in terms of QRPA phonons given as
\begin{equation}\label{co}
	A^{\dagger}_{\omega}(\mu)=\sum_{pn}[X^{pn}_{\omega}(\mu)a^{\dagger}_{p}a^{\dagger}_{\overline{n}}-Y^{pn}_{\omega}(\mu)a_{n}a_{\overline{p}}],
\end{equation}
where $p$ ($n$) stands for $m_{p}\alpha_{p}$ ($m_{n}\alpha_{n}$). The sum includes all proton-neutron pairs subject to conditions $\mu=m_{p}-m_{n}$ and $\pi_{p}.\pi_{n}$=1, with $\pi$ denoting parity. The $X$ ($Y$) denotes the forward-going (backward-going) amplitude. The proton-neutron residual interactions in pn-QRPA approach occur {via separable \textit{pp} and \textit{ph} channels, which are defined by interaction constants $\kappa$ and $\chi$, respectively.} The $ph$ GT force is denoted as
\begin{equation}\label{ph}
	V^{ph}= +2\chi\sum^{1}_{\mu= -1}(-1)^{\mu}Y_{\mu}Y^{\dagger}_{-\mu},\\
\end{equation}
with
\begin{equation}\label{y}
	Y_{\mu}= \sum_{j_{p}m_{p}j_{n}m_{n}}<j_{p}m_{p}\mid
	t_- ~\sigma_{\mu}\mid
	j_{n}m_{n}>c^{\dagger}_{j_{p}m_{p}}c_{j_{n}m_{n}},
\end{equation}
and the $pp$ GT force as
\begin{equation}\label{pp}
	V^{pp}= -2\kappa\sum^{1}_{\mu=
		-1}(-1)^{\mu}P^{\dagger}_{\mu}P_{-\mu},
\end{equation}
with
\begin{eqnarray}\label{p}
	P^{\dagger}_{\mu}= \sum_{j_{p}m_{p}j_{n}m_{n}}<j_{n}m_{n}\mid
	(t_- \sigma_{\mu})^{\dagger}\mid
	j_{p}m_{p}>\times \nonumber\\
	(-1)^{l_{n}+j_{n}-m_{n}}c^{\dagger}_{j_{p}m_{p}}c^{\dagger}_{j_{n}-m_{n}},
\end{eqnarray}
where the rest of the symbols have their traditional meanings. The \textit{ph} and \textit{pp} force have different signs revealing their opposite nature. The interaction strengths $\kappa$ and $\chi$ were chosen in accordance with Ref.~\cite{Hom96}, based on a $1/A^{0.7}$ relationship. The interaction strengths (\textit{ph} and \textit{pp}) were included consistently for both $\beta^+$ and $\beta^-$ decays, and their values were fixed as smooth functions of nuclei' mass number A in such a way that the computation best mirrored the known $\beta$-decay features. Our calculation also satisfied the model independent Ikeda sum rule \cite{Ike63}. The reduced GT transition probabilities from the QRPA ground state to one-phonon states in the daughter nucleus were calculated as
\begin{equation}
	B_{GT} (\omega) = |\langle \omega, \mu ||t_{+} \sigma_{\mu}||QRPA \rangle|^2.
\end{equation}
We refer to Ref.~\cite{Hir93} for more information and for diagonalization of the Hamiltonian (Eq.~(\ref{H})). \\
%The deformation parameter ($\beta_2$) was calculated using $\beta_2$ = $\frac{125 Q_2}{1.44ZA^{2/3}}$ where the quadrupole moment $Q_2$ was taken from Ref.~\cite{Mol81}.
The $\beta$-decay partial half-lives were calculated employing
\begin{eqnarray}
	t_{p(1/2)} = \nonumber\\	\frac{C}{(g_A/g_V)^2f_A(Z, A, E)B_{GT}(E_j)+f_V(Z, A, E)B_F(E_j)},
\end{eqnarray}
where $E$ = $Q$ - $E_j$, $E_j$ and $g_A/g_V$ (= -1.254)\cite{War94} denote energy of the final state and ratio of axial to the vector coupling constant, respectively and C = ${2\pi^3 \hbar^7 ln2}/{g^2_V m^5_ec^4} = 6295 s$. $f_A(E, Z, A)$ and $f_V(E, Z, A)$ are the phase space integrals for axial vector and vector transitions, respectively. $B_{F}$ ($B_{GT}$) stands for the reduced transition probability for the Fermi (GT) transitions.
Finally, the total $\beta$-decay half-lives were calculated using the equation
\begin{equation}
	T_{1/2} = \left(\sum_{0 \le E_j \le Q} \frac{1}{t_{p(1/2)}}\right)^{-1}.
\end{equation} 
The summation includes all the transition probabilities to the states in daughter within the $Q$ window.\\
As mentioned earlier, pairing gap values are key model parameters in the pn-QRPA approach. Two different values of pairing gaps were used in the current calculation. The first one was computed using the traditional and  mass-dependant relation $\vartriangle_p=\vartriangle_n={12/\sqrt A}$ MeV~\cite{hardy09}. The second scheme, which consists of three terms, computes different pairing gaps for neutrons and protons and is expressed in terms of proton and neutron separation energies as:
\begin{eqnarray}
	\bigtriangleup_{pp} =\frac{1}{4}(-1)^{Z+1}[S_p(Z+1, A+1)\nonumber\\~~~~~~~~~~~~~~~~~~~~~~-2S_p(Z, A)+S_p(Z-1, A-1)]
\end{eqnarray}
\begin{eqnarray}
	\bigtriangleup_{nn} =\frac{1}{4}(-1)^{A-Z+1}[S_n(Z, A+1)\nonumber\\~~~~~~~~~~~~~~~~~~~~~~~~~~~~- 2S_n(Z, A) + S_n(Z, A-1)]
\end{eqnarray} 
For convenience, we refer to the first scheme as TF and the second scheme as 3TF, respectively. 
%%%%%%%%%%%%%%%%%%%%%%%%%%%%%%% 
\section{Results and Discussion} 
Reliable estimates of the $\beta$-decay properties of neutron-rich nuclei are necessary  to better understand supernova explosions and associated nucleosynthesis processes, particularly the $r$-process. In this work we focus on the computation of $\beta$-decay half-lives and branching ratios of the GT strength distributions of  neutron-rich Cerium isotopes using the pn-QRPA approach.  Among the 34 isotopes of Ce considered in the current calcuation, 16 decay via positron emission and 16 (2) nuclei are unstable to electron emission (electron capture). $^{136, 138, 140, 142}$Ce are stable isotopes. We employed two different values of the pairing gaps (TF and 3TF) in our computation. The computed decay half-lives of the considered cases are compared with measured data adopted from Kondev et al. \cite{Aud21}.\\
{On the basis of computed pairing gap values, we may face four different cases. The 	$\bigtriangleup_{pp}$ values, calculated using the 3TF scheme, can be bigger (or smaller) as compared to the traditional choice of  $\bigtriangleup_{pp=nn}$ values of the TF scheme \textit{and} the 	$\bigtriangleup_{nn}$ values, calculated using the 3TF scheme, can be smaller (or bigger) as compared to the  $\bigtriangleup_{pp=nn}$ of the TF scheme. These two cases would henceforth be denoted by C1 and C2, respectively. On the other hand, it is also possible that \textit{both} $\bigtriangleup_{pp}$ and $\bigtriangleup_{nn}$ values of the 3TF scheme are smaller or bigger than the $\bigtriangleup_{pp=nn}$ values of the TF scheme. The latter two cases would be referred to as C3 and C4, respectively, in this paper. It is to be noted that we considered an equal number of $\beta^+$ and $\beta^-$ cases for the current investigation. }\\
The sample GT strength distributions for the case  of $^{125}$Ce ($\beta^+$ decay) and $^{146}$Ce ($\beta^-$ deacy), using the TF and 3TF computed pairing gaps, are shown in Fig.~1. Fragmentation of GT strength is more pronounced in the TF scheme. The figure clearly shows that the two schemes produce distinct strength distributions. This change in the strength distributions further affects the computed half-lives and centroids, which we will discuss later. It is worth mentioning that we show the GT strength lying within the $Q$-window only. \\
Table~1 shows the pairing gap and half-life values calculated employing the two different schemes (TF and 3TF), {for the C1, C2 and C3 cases, while Table~2 presents similar data for the C4 cases. The trend in the computed half-lives may be explained by investigating the cumulative GT strength and centroid values which we discuss later. We further compare} the calculated half-life values to the experimental data \cite{Aud21} (displayed in the last column of Tables~1 \& 2). It is noted from Table~1 and Table~2 that there exists up to two orders of magnitude difference between calculated and measured half-lives of odd-A isotopes  $^{139, 141, 143}$Ce. This may be attributed to the N=82 shell closure effect. All the even-even isotopes in the vicinity,  $^{136, 138, 140, 142}$Ce, are stable and nearby odd-A isotopes exhibit larger deviations with the measured data.  
%It should be noted that the two schemes produce more or less similar half-life values for few $\beta^+$ cases such as $^{123}$Ce, $^{127}$Ce, $^{130-133}$Ce and $^{137}$Ce, an \textit{ec} case $^{139}$Ce and a few $\beta^-$ cases like $^{141}$Ce, $^{143}$Ce, $^{145}$Ce, $^{147}$Ce, $^{151-152}$Ce and $^{154}$Ce. However, for the cases like $^{120}$Ce, $^{128}$Ce, $^{135}$Ce, $^{146}$Ce and $^{149}$Ce, substantial differences in the computed half-lives were noted. 
The 3TF scheme yields calculated half-lives that are in better agreement with the measured data. The half-lives computed employing the 3TF scheme are within a factor 2 (5) for 20 (6) cases, while 8 cases fall beyond factor 5 difference when compared with the measured data. However, in the case of TF scheme, a total of 12 cases fall beyond factor 5 difference in comparison with the measured half-lives. {As mentioned earlier, the} computed centroids and GT strength distributions, which we address next, are likely reasons for the variations in the computed half-lives when different pairing gap values are used.\\ 
{The computed total GT strength (in arbitrary units) and centroid values (in MeV units) of the pn-QRPA calculated GT strength distributions for the C1, C2 and C3 (C4) cases are presented in Table 3 (Table 4). We note that there are cases (e.g. $^{120-122, 133, 134, 139, 141, 145, 149}$Ce) where the total strength values vary significantly as the pairing gap values change. The centroid values shifted as well with changing pairing gap values. It is noted that, in general, the TF scheme resulted in smaller total GT strength values for the C1 and C4 cases (see Table 3 and Table 4). The smaller total strength values resulted in lower rates and correspondingly bigger calculated half-lives (see Table 1 and Table 2). The, in general, bigger computed centroid values using the 3TF scheme led to smaller rates and correspondingly bigger calculated half-lives for the C2 and C3 cases (see Table 1). For C1, C2 and C4 cases, the differences in computed total GT strength and centroid values using the two schemes resulted up to  an order of magnitude difference in the calculated half-lives. Only in the C3 case are the computed half-lives, using the TF and 3TF schemes, in decent agreement with each other. We would like to remain cautious in correlating the differences in calculated half-lives with the four cases. This warrants a plentiful of decay cases which we would like to take as a future assignment. } 
%When compared to the TF scheme, the higher total strength value of $^{120}$Ce employing 3TF scheme resulted in a higher weak rate and a correspondingly lower calculated half-life (see Table 1). Similarly, the lower strength values of $^{134, 146}$Ce using TF resulted in bigger calculated half-lives. Another factor, that influences the calculated half-lives and centroid values, is the re-distribution of GT strength. The GT strength occasionally shifts to  higher excitation energies beyond the $Q$-window. The TF method locates the centroid value at a considerably lower excitation energy than 3TF for $^{129}$Ce. The low centroid value resulted in a higher rate and, as a result, a shorter computed half-life (Table 1). Other cases showed similar trends.
\\   
The branching ratio was calculated employing the following relation:
\begin{eqnarray}
	I = \frac{T_{1/2}}{t^{par}_{1/2}} \times 100 (\%)
\end{eqnarray}
where $T_{1/2}$ stands for the total half-life.
Table 5 shows the GT strength (state-by-state), branching ratio (denoted by I), and partial half-lives ($t^{par}_{1/2}$) for the decay of $^{146}$Ce. It is noted that the TF scheme resulted in more fragmentation of the GT strength (see also Fig.~1). The 3TF scheme computed high branching ratios ($\sim$21, $\sim$33 \& $\sim$11) at low excitation energies ($\sim$0.15, $\sim$0.19 \& $\sim$0.33, respectively) resulting in low centroid value of 0.49 versus 0.72 of the TF scheme. The smaller centroid value led to a higher rate and, as a result, a shorter half-life. Furthermore, the TF scheme calculated a lower value of the total strength which resulted in smaller rate value and, as a result, a high value of calculated half-life. State-by-state GT values, branching ratios and partial half-lives for remaining 33 isotopes of Ce were also calculated and may be requested as ASCII files from the corresponding author.\\

%%%%%%%%%%%%%%%%%%%%%%%%%%%%%%%%%%%%%%%%%%%%%%%%%%%%%%%%%%%%%%%%%%%%%%%%%%%%%%%%%%%%%%%%%%%%%%%%%%%%%%%%
\section{Summary and Conclusion} 
We report the computation of $\beta$-decay properties of the neutron-rich Cerium isotopes. The $\beta$-decay half-lives and branching ratios for a total of 34 Ce isotopes in the mass range 120 $\le$ A $\le$ 157 were calculated using the pn-QRPA model. Pairing gap between the paired nucleons is one of the key model parameters in the pn-QRPA theory. In this work we investigate the effects of altering pairng gap values on the computed $\beta$-decay properties of the neutron-rich Cerium isotopes.  Two different values of pairing gap were used in our calculation. The GT strength distributions and centroid values were significantly altered as the pairing gap values were changed. Consequently, the $\beta$-decay half-life values also changed. The T$_{1/2}$ values calculated using the three term formula (3TF), based on neutron and proton separation energies, were found to match {better} the measured data. {In future we would like to include hundreds of unstable nuclei in our pool to study the possible correlation of pairing gap values with calculated $\beta$-decay half-lives. We would further like to investigate the improvement in current calculation by using a realistic nucleon-nucleon potential in our Hamiltonian.  }

%%%%%%%%%%%%%%%%%%%%%% Acknowledgment %%%%%%%%%%%%%%%%%%%%%%%%%%
%
%\vspace{0.5in} \textbf{Acknowledgment}:  
%N. Cakmak would like to thank C. Selam for very fruitful
%discussion on calculation of GT and FF transitions.

\newpage
%%%%%%%%%%%%%%%%%%%%%%%%%%% References %%%%%%%%%%%%%%%%%%%%%%%%%%%%%

\newpage
\begin{figure*}
	\hfil
	\vspace*{2.5cm}
	\hspace*{-1.5cm}	\includegraphics*[width=0.9\paperwidth]{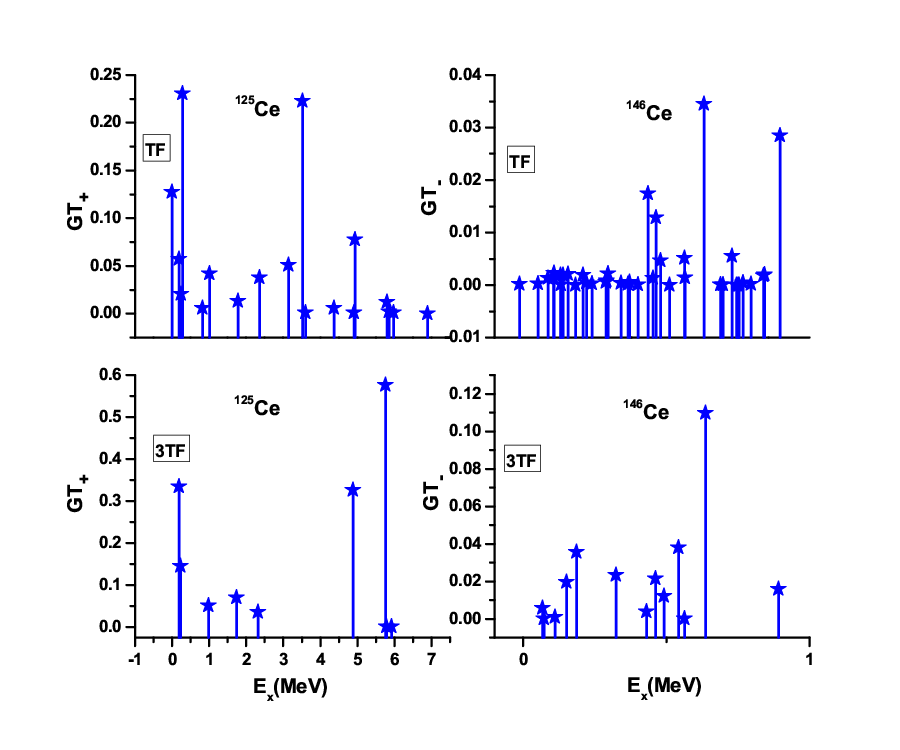}
	\vspace*{-1.5cm}
	 \caption{Calculated GT strength distributions for $^{125}$Ce and $^{146}$Ce as a function of excitation energy in daughter.}
\end{figure*}\label{F1}

%TABLES
%%%%%%%%%%%%%%%%%%%%%%%%%%%%%%%%%%%%%%%%%%%%%
\begin{table*}[]\label{T1}
	\centering
	\scriptsize \caption{Computed pairing gaps and half-life values {for the C1 (upper panel), C2 (middle panel) and C3 (bottom panel) cases} in comparison with the experimental data \cite{Aud21}.  }
	
	\begin{tabular}{c|c|ccc|cc|c}
		\hline
		&&\multicolumn{3}{c}{Pairing Gaps (MeV)}&\multicolumn{3}{c}{Half-lives (s)}\\
		\hline
		Nuclei&Decay Mode&$\bigtriangleup^{TF}_{nn=pp}$&$\bigtriangleup^{3TF}_{nn}$&$\bigtriangleup^{3TF}_{pp}$&$T^{TF}_{1/2}$&$T^{3TF}_{1/2}$&$T^{Exp}_{1/2}$ \\
		\hline
		$^{139}$Ce & \textit{ec} 	  & 1.02 & 1.00 & 1.09 & 3.42E+05 & 3.18E+05 & 1.19E+07 \\
		$^{144}$Ce & $\beta^-$& 1.00 & 0.98 & 1.25 & 1.76E+06 & 1.01E+07 & 2.46E+07 \\
		$^{150}$Ce & $\beta^-$& 0.98 & 0.93 & 1.16 & 1.12E+01 & 1.28E+01 & 6.05E+00 \\
		$^{152}$Ce & $\beta^-$& 0.97 & 0.80 & 1.33 & 2.06E+00 & 1.94E+00 & 1.42E+00 \\
		$^{153}$Ce & $\beta^-$& 0.97 & 0.80 & 0.99 & 6.09E-01 & 8.39E-01 & 8.65E-01 \\
		$^{154}$Ce & $\beta^-$& 0.97 & 0.78 & 1.17 & 9.54E-01 & 9.00E-01 & 7.22E-01 \\
		$^{156}$Ce & $\beta^-$& 0.96 & 0.85 & 1.29 & 3.56E-01 & 1.86E-01 & 2.33E-01 \\
		$^{157}$Ce & $\beta^-$& 0.96 & 0.86 & 5.87 & 3.51E-01 & 2.38E-01 & 1.75E-01 \\
		\hline
		$^{131}$Ce & $\beta^+$& 1.05 & 1.33 & 1.02 & 4.11E+01 & 4.16E+01 & 6.18E+02 \\
		$^{145}$Ce & $\beta^-$& 1.00 & 1.03 & 0.98 & 1.85E+02 & 1.75E+02 & 1.81E+02 \\
		$^{147}$Ce & $\beta^-$& 0.99 & 1.04 & 0.84 & 2.94E+01 & 3.15E+01 & 5.64E+01 \\
		$^{149}$Ce & $\beta^-$& 0.98 & 1.00 & 0.96 & 6.57E+00 & 2.36E+01 & 4.94E+00 \\
		\hline
		$^{124}$Ce & $\beta^+$& 1.08 & 0.80 & 0.86 & 1.20E+01 & 1.21E+01 & 9.10E+00 \\
		$^{143}$Ce & $\beta^-$& 1.00 & 0.95 & 0.93 & 1.52E+03 & 1.61E+03 & 1.19E+05 \\
		$^{151}$Ce & $\beta^-$& 0.98 & 0.79 & 0.95 & 1.78E+00 & 1.73E+00 & 1.76E+00 \\
		$^{155}$Ce & $\beta^-$& 0.96 & 0.81 & 0.94 & 3.32E+00 & 3.42E+00 & 3.13E-01 \\
		\hline
	\end{tabular}
\end{table*}
\begin{table*}[]\label{T2}
	\centering
	\scriptsize \caption{Same as Table~1, but for {the C4 cases.}  }
	
	\begin{tabular}{c|c|ccc|cc|c}
		\hline
		&&\multicolumn{3}{c}{Pairing Gaps (MeV)}&\multicolumn{3}{c}{Half-lives (s)}\\
		\hline
		Nuclei&Decay Mode&$\bigtriangleup^{TF}_{nn=pp}$&$\bigtriangleup^{3TF}_{nn}$&$\bigtriangleup^{3TF}_{pp}$&$T^{TF}_{1/2}$&$T^{3TF}_{1/2}$&$T^{Exp}_{1/2}$ \\
		\hline
		$^{120}$Ce & $\beta^+$& 1.10 & 8.73 & 1.20 & 1.80E+00 & 9.61E-01 & 2.50E-01 \\
		$^{121}$Ce & $\beta^+$& 1.09 & 1.29 & 1.29 & 7.58E-01 & 1.14E+00 & 1.10E+00 \\
		$^{122}$Ce & $\beta^+$& 1.09 & 1.24 & 1.43 & 5.57E+00 & 2.98E+00 & 2.00E+00 \\
		$^{123}$Ce & $\beta^+$& 1.08 & 1.24 & 1.21 & 1.98E+00 & 2.01E+00 & 3.80E+00 \\
		$^{125}$Ce & $\beta^+$& 1.07 & 1.33 & 1.13 & 1.93E+00 & 2.37E+00 & 9.70E+00 \\
		$^{126}$Ce & $\beta^+$& 1.07 & 1.20 & 1.43 & 9.12E+01 & 6.21E+01 & 5.10E+01 \\
		$^{127}$Ce & $\beta^+$& 1.06 & 1.35 & 1.09 & 2.15E+01 & 2.13E+01 & 3.40E+01 \\
		$^{128}$Ce & $\beta^+$& 1.06 & 1.30 & 1.45 & 3.60E+02 & 2.27E+02 & 2.36E+02 \\
		$^{129}$Ce & $\beta^+$& 1.06 & 1.30 & 1.15 & 7.47E+01 & 1.32E+02 & 2.10E+02 \\
		$^{130}$Ce & $\beta^+$& 1.05 & 1.31 & 1.34 & 1.48E+03 & 1.37E+03 & 1.37E+03 \\
		$^{132}$Ce & $\beta^+$& 1.04 & 1.32 & 1.35 & 9.20E+03 & 1.36E+04 & 1.26E+04 \\
		$^{133}$Ce & $\beta^+$& 1.04 & 1.32 & 1.06 & 1.02E+02 & 1.01E+02 & 5.82E+03 \\
		$^{134}$Ce & \textit{ec} 	& 1.04 & 1.27 & 1.38 & 2.94E+05 & 1.85E+05 & 2.73E+05 \\
		$^{135}$Ce & $\beta^+$& 1.03 & 1.19 & 1.10 & 4.38E+04 & 5.68E+03 & 6.37E+04 \\
		$^{137}$Ce & $\beta^+$& 1.03 & 1.18 & 1.09 & 1.75E+03 & 2.05E+03 & 3.24E+04 \\
		$^{141}$Ce & $\beta^-$& 1.01 & 1.38 & 1.12 & 3.40E+04 & 3.21E+04 & 2.81E+06 \\
		$^{146}$Ce & $\beta^-$& 0.99 & 1.02 & 1.17 & 6.05E+03 & 1.56E+03 & 8.09E+02 \\
		$^{148}$Ce & $\beta^-$& 0.99 & 1.03 & 1.25 & 1.51E+02 & 1.40E+02 & 5.68E+01 \\
		\hline
	\end{tabular}
\end{table*}

\begin{table*}[]\label{T3}
	\centering
	\scriptsize \caption{Calculated centroid and total GT strength values of the GT strength distributions {for the C1 (upper panel), C2 (middle panel) and C3 (bottom panel) cases} using the two computed pairing gaps (TF and 3TF).}
	\centering
	\begin{tabular}{c|c|cc|cc}
		\hline
		& 	&\multicolumn{2}{c|}{Total Strength (arb. units)}&\multicolumn{2}{c}{Centroid (MeV)}\\
		\hline
		Nuclei& Q-value (MeV)&$\sum$ GT$^{TF}$ &$\sum$ GT$^{3TF}$ &$\bar{E}$$^{TF}$&$\bar{E}$$^{3TF}$\\
		\hline
		$^{139}$Ce & 0.26 & 0.06 & 0.46 & 0.04 & 0.03  \\
		$^{144}$Ce & 0.32 & 0.06 & 0.01 & 0.21 & 0.22  \\
		$^{150}$Ce & 3.45 & 0.47 & 0.81 & 1.81 & 2.34  \\
		$^{152}$Ce & 4.78 & 1.39 & 1.08 & 2.13 & 3.05  \\
		$^{153}$Ce & 6.66 & 0.39 & 0.89 & 3.53 & 4.48  \\
		$^{154}$Ce & 5.64 & 1.97 & 2.03 & 1.40 & 1.17  \\
		$^{156}$Ce & 6.63 & 2.77 & 2.34 & 1.14 & 0.98  \\
		$^{157}$Ce & 8.51 & 1.96 & 2.39 & 6.54 & 6.17  \\
		\hline
		$^{131}$Ce & 4.06 & 0.56 & 0.55 & 0.50 & 0.51  \\
		$^{145}$Ce & 2.56 & 0.17 & 0.04 & 0.93 & 0.89  \\
		$^{147}$Ce & 3.43 & 0.11 & 0.11 & 1.32 & 1.37  \\
		$^{149}$Ce & 4.37 & 0.21 & 1.07 & 2.41 & 3.26  \\
		\hline 
		$^{124}$Ce & 5.97 & 2.35 & 2.35 & 3.78 & 3.85  \\
		$^{143}$Ce & 1.46 & 0.04 & 0.08 & 0.41 & 0.49  \\
		$^{151}$Ce & 5.56 & 0.22 & 0.23 & 2.92 & 2.93  \\
		$^{155}$Ce & 7.64 & 0.22 & 0.24 & 4.30 & 4.32  \\
		\hline
	\end{tabular}
\end{table*}
\begin{table*}[]\label{T4}
	\centering
	\scriptsize \caption{Same as Table~3, but for {the C4 cases.}}
	\centering
	\begin{tabular}{c|c|cc|cc}
		\hline
		& 	&\multicolumn{2}{c|}{Total Strength (arb. units)}&\multicolumn{2}{c}{Centroid (MeV)}\\
		\hline
		Nuclei&Q-value (MeV) &$\sum$ GT$^{TF}$ &$\sum$ GT$^{3TF}$ &$\bar{E}$$^{TF}$&$\bar{E}$$^{3TF}$\\
		\hline
		$^{120}$Ce & 7.84 & 3.42 & 8.93 & 1.66 & 4.24  \\
		$^{121}$Ce & 9.50 & 0.57 & 7.34 & 1.73 & 6.73  \\
		$^{122}$Ce & 6.67 & 2.73 & 6.92 & 1.13 & 3.62  \\
		$^{123}$Ce & 8.36 & 6.57 & 6.33 & 6.11 & 6.29  \\
		$^{125}$Ce & 7.10 & 0.91 & 1.54 & 1.86 & 3.42  \\
		$^{126}$Ce & 4.15 & 1.29 & 3.83 & 2.46 & 2.82  \\
		$^{127}$Ce & 5.92 & 1.62 & 0.41 & 3.52 & 3.70  \\
		$^{128}$Ce & 3.10 & 0.91 & 2.54 & 1.90 & 2.12  \\
		$^{129}$Ce & 5.04 & 2.62 & 2.43 & 1.80 & 4.26  \\
		$^{130}$Ce & 2.20 & 0.65 & 0.55 & 1.48 & 1.32  \\
		$^{132}$Ce & 1.25 & 0.18 & 0.12 & 0.52 & 0.60  \\
		$^{133}$Ce & 3.08 & 0.50 & 0.08 & 0.04 & 0.04  \\
		$^{134}$Ce & 0.39 & 0.04 & 0.15 & 0.25 & 0.23  \\
		$^{135}$Ce & 2.03 & 0.05 & 0.08 & 1.51 & 0.36  \\
		$^{137}$Ce & 1.22 & 0.39 & 0.40 & 0.13 & 0.20  \\
		$^{141}$Ce & 0.58 & 0.03 & 0.16 & 0.22 & 0.33  \\
		$^{146}$Ce & 1.04 & 0.13 & 0.29 & 0.72 & 0.49  \\
		$^{148}$Ce & 2.14 & 0.32 & 0.38 & 1.22 & 1.25  \\
		\hline
	\end{tabular}
\end{table*}

\begin{table*}[]\label{T6}
	\centering
	\scriptsize\caption{The state by state GT strength, partial half-life (in unit of $s$) values and branching ratios (I) for $^{146}$Ce using the two pairing gaps (TF and 3TF).}
	\begin{tabular}{cccc|cccc}
		\hline
		\multicolumn{4}{c}{TF} & \multicolumn{4}{c}{3TF} \\
		\hline
		E$_x$ (MeV)& GT & I&t$^{par}_{1/2}$&E$_x$ (MeV)& GT & I&t$^{par}_{1/2}$ \\
		\hline
		0.091 & 0.00020 & 0.02  & 4.38E+07 & 0.004 & 0.00074 & 1.40  & 1.12E+05 \\
		0.159 & 0.00030 & 1.87  & 4.99E+05 & 0.068 & 0.00578 & 8.66  & 1.80E+04 \\
		0.195 & 0.00133 & 7.17  & 1.30E+05 & 0.073 & 0.00025 & 0.36  & 4.31E+05 \\
		0.215 & 0.00136 & 6.71  & 1.39E+05 & 0.111 & 0.00085 & 1.08  & 1.45E+05 \\
		0.218 & 0.00233 & 11.45 & 8.14E+04 & 0.151 & 0.01952 & 21.20 & 7.37E+03 \\
		0.241 & 0.00193 & 8.55  & 1.09E+05 & 0.186 & 0.03548 & 33.34 & 4.69E+03 \\
		0.242 & 0.00011 & 0.46  & 2.01E+06 & 0.325 & 0.02338 & 11.74 & 1.33E+04 \\
		0.250 & 0.00189 & 8.02  & 1.16E+05 & 0.431 & 0.00384 & 1.11  & 1.41E+05 \\
		0.269 & 0.00214 & 8.36  & 1.11E+05 & 0.463 & 0.02149 & 5.15  & 3.03E+04 \\
		0.295 & 0.00002 & 0.08  & 1.13E+07 & 0.492 & 0.01213 & 2.45  & 6.39E+04 \\
		0.323 & 0.00194 & 5.87  & 1.59E+05 & 0.543 & 0.03809 & 5.54  & 2.82E+04 \\
		0.335 & 0.00046 & 1.30  & 7.16E+05 & 0.563 & 0.00007 & 0.01  & 1.84E+07 \\
		0.356 & 0.00026 & 0.66  & 1.41E+06 & 0.637 & 0.10969 & 7.91  & 1.98E+04 \\
		0.407 & 0.00057 & 1.11  & 8.40E+05 & 0.891 & 0.01577 & 0.05  & 3.14E+06 \\
		0.408 & 0.00089 & 1.74  & 5.37E+05 &       &         &       &          \\
		0.414 & 0.00215 & 4.06  & 2.29E+05 &       &         &       &          \\
		0.462 & 0.00036 & 0.52  & 1.79E+06 &       &         &       &          \\
		0.487 & 0.00021 & 0.26  & 3.54E+06 &       &         &       &          \\
		0.492 & 0.00056 & 0.67  & 1.40E+06 &       &         &       &          \\
		0.523 & 0.00005 & 0.05  & 1.73E+07 &       &         &       &          \\
		0.560 & 0.01745 & 13.37 & 6.96E+04 &       &         &       &          \\
		0.578 & 0.00132 & 0.91  & 1.03E+06 &       &         &       &          \\
		0.589 & 0.01288 & 8.07  & 1.15E+05 &       &         &       &          \\
		0.606 & 0.00462 & 2.54  & 3.67E+05 &       &         &       &          \\
		0.638 & 0.00001 & 0.00  & 2.24E+08 &       &         &       &          \\
		0.694 & 0.00511 & 1.34  & 6.96E+05 &       &         &       &          \\
		0.696 & 0.00144 & 0.37  & 2.51E+06 &       &         &       &          \\
		0.765 & 0.03445 & 4.31  & 2.16E+05 &       &         &       &          \\
		0.825 & 0.00009 & 0.01  & 1.75E+08 &       &         &       &          \\
		0.834 & 0.00011 & 0.01  & 1.65E+08 &       &         &       &          \\
		0.867 & 0.00549 & 0.16  & 5.71E+06 &       &         &       &          \\
		0.884 & 0.00000 & 0.00  & 2.72E+10 &       &         &       &          \\
		0.889 & 0.00000 & 0.00  & 1.32E+10 &       &         &       &          \\
		0.893 & 0.00002 & 0.00  & 2.39E+09 &       &         &       &          \\
		0.908 & 0.00052 & 0.01  & 1.36E+08 &       &         &       &          \\
		0.936 & 0.00021 & 0.00  & 6.94E+08 &       &         &       &          \\
		0.982 & 0.00183 & 0.00  & 4.41E+08 &       &         &       &          \\
		0.986 & 0.00198 & 0.00  & 4.82E+08 &       &         &       &          \\
		1.042 & 0.02843 & 0.00  & 4.85E+11 &       &         &       &   \\		
		\hline
	\end{tabular}
\end{table*}
\end{document}